\documentclass[twocolumn,aps,amsiath,amssymb,showkeys,floatfix,pra,superscriptaddress]{revtex4-1}

\usepackage{bbm}
\usepackage[english]{babel}
\usepackage[dvips]{graphicx}
\usepackage{color}
\usepackage{amsmath,amssymb,latexsym}

\newcommand{\R}{\mathbf{R}}

\begin{document}

\title{Optimal decision making for sperm chemotaxis in the presence of noise}

\author{Justus A. Kromer}
\affiliation{cfaed, TU Dresden, Dresden, Germany}
\author{Steffen M\"arcker}
\affiliation{Faculty of Computer Science, TU Dresden, Dresden, Germany}
\author{Steffen Lange} 
\affiliation{cfaed, TU Dresden, Dresden, Germany}
\author{Christel Baier}
\affiliation{Faculty of Computer Science, TU Dresden, Dresden, Germany}
\author{Benjamin M. Friedrich}
\affiliation{cfaed, TU Dresden, Dresden, Germany}

\date{\today}


\begin{abstract}
For navigation, microscopic agents such as biological cells rely on noisy sensory input. In cells performing chemotaxis, such noise arises from the stochastic binding of signaling molecules at low concentrations. Using chemotaxis of sperm cells as application example, we address the classic problem of chemotaxis towards a single target. We reveal a fundamental relationship between the speed of chemotactic steering and the strength of directional fluctuations that result from the amplification of noise in the chemical input signal. This relation implies a trade-off between slow, but reliable, and fast, but less reliable, steering.

By formulating the problem of optimal navigation in the presence of noise as a Markov decision process, we show that dynamic switching between reliable and fast steering substantially increases the probability to find a target, such as the egg. Intriguingly, this decision making would provide no benefit in the absence of noise. Instead, decision making is most beneficial, if chemical signals are above detection threshold, yet signal-to-noise ratios of gradient measurements are low. This situation generically arises at intermediate distances from a target, where signaling molecules emitted by the target are diluted, thus defining a `noise zone' that cells have to cross.

Our work addresses the intermediate case between well-studied perfect chemotaxis at high signal-to-noise ratios close to a target, and random search strategies in the absence of navigation cues, e.g. far away from a target. Our specific results provide a rational for the surprising observation of decision making in recent experiments on sea urchin sperm chemotaxis. The general theory demonstrates how decision making enables chemotactic agents to cope with high levels of noise in gradient measurements by dynamically adjusting the persistence length of a biased persistent random walk.
\end{abstract}

\maketitle

\section{Introduction}

Motile cells successfully navigate in external concentration fields of signaling molecules
by steering in the direction of local concentration gradients -- a process termed chemotaxis.
Chemotaxis represents a biological implementation of a gradient-ascent algorithm and is used by 
bacteria to find food \cite{sourjik2012responding},
immune cells to locate infection sites \cite{Devreotes1988}, and
sperm cells to follow gradients of chemical cues to find the egg \cite{Eisenbach2006,Alvarez2014}. 
The very task of reliably measuring a 
local concentration gradient with sufficient accuracy is non-trivial at dilute concentrations, 
since molecular shot noise corrupts concentration measurements 
\cite{berg1977physics,bialek2005physical,endres2008accuracy,hein2016physical}. 
To measure a concentration, cells must count individual binding events of signaling molecules, 
which represents a stochastic Poisson process.

Pioneering work on this topic studied the chemotaxis of mobile agents with advanced information processing skills \cite{gomez2011}, 
or even a capacity to compute spatial maps of maximum likelihood of target position~\cite{Vergassola2007}.
It is an open question how biological cells with limited information processing capability 
deal with noise during their chemotaxis~\cite{Andrews2006,Celani2010,Amselem2012}. 
This question prompts a combined analysis of both the physics of cellular navigation and 
the signal-to-noise ratios that these cells typically encounter in their environment.

Here, we present a theory of optimal chemotaxis strategies in the presence of noise,
using the framework of Markov decision processes (MDP).
This general approach opens a new route to study optimal navigation strategies. 

We apply this theoretical approach to chemotaxis along helical swimming paths,
which is employed by sperm cells of marine species.
Chemotaxis along helical paths represents one of the three fundamental gradient-sensing strategies of biological cells \cite{Alvarez2014}. 
It is based on 
temporal comparison of a concentration signal along the swimming path \cite{Crenshaw1990,Friedrich2007,Jikeli2015}. 
By swimming along a helical path, {i.e.} circling around a centerline, 
these cells receive information about the gradient component perpendicular to their direction of net motion. 
This information enables cells to steer in a directed manner, 
by bending the direction of their helical paths towards the local gradient,
see also Fig.~\ref{fig:exp}A. 
Helical swimming represents a stereotypical form of exploratory behavior, 
employed by sperm cells and other microswimmers \cite{Witman1993}. 
This strategy is typical for sperm cells from species with external fertilization 
\cite{Brokaw1958,Miller1985,Jikeli2015}.

The model system of sperm chemotaxis is particularly suited 
to address optimal navigation in the presence of noise:
First, sperm cells have a single objective, to find the egg.
In species with external fertilization, 
evolution presumably optimized the probability to find an egg. 
Second, recent experiments revealed that sea urchin sperm cells
dynamically switch between two different steering modes \cite{Jikeli2015},
thus providing an instance of cellular decision making.
To date, the benefit of this decision making is not known.
With our theory, we demonstrate a benefit of decision making in sperm chemotaxis, 
and show that this benefit is directly related to noise in cellular gradient sensing.

Our work addresses the intermediate case between 
the well-understood case of perfect chemotaxis in the absence of noise (perfectly reliable steering), 
and purely random search strategies that operate in the absence of directed signals (no steering)
\cite{Benichou2011,Hein2012,Bartumeus2005,Friedrich2008}.
We show that even if the signal-to-noise ratio of gradient-sensing is below one, 
thus impeding reliable chemotactic steering,
situation-specific switching between two steering modes 
can substantially increase the probability to find a target, such as the egg.
This applies in particular at intermediate distances from the target, in a `noise zone', 
which cells have to cross before they can perform perfect chemotaxis close to the target.

Sperm chemotaxis along helical paths represents a prototype for cellular motility with directional persistence.
The centerline of helical swimming paths follows a biased persistent random walk,
whose tangent vector gradually aligns with the local concentration gradient. 
Our theory highlights a fundamental relationship between 
the speed of this chemotactic re-orientation and the strength of directional fluctuations,
which result from the amplification of noise in the chemotactic input signal. 
Thus, responding faster comes at the price of increased stochasticity of swimming paths and decreased directional persistence. 
This implies a trade-off choice between speed and reliability in navigation,
for which decision making represents an efficient solution with minimal control complexity.


\section{Decision making in sperm chemotaxis: previous experiments}

\begin{figure}[t]
\includegraphics[width=1\linewidth]{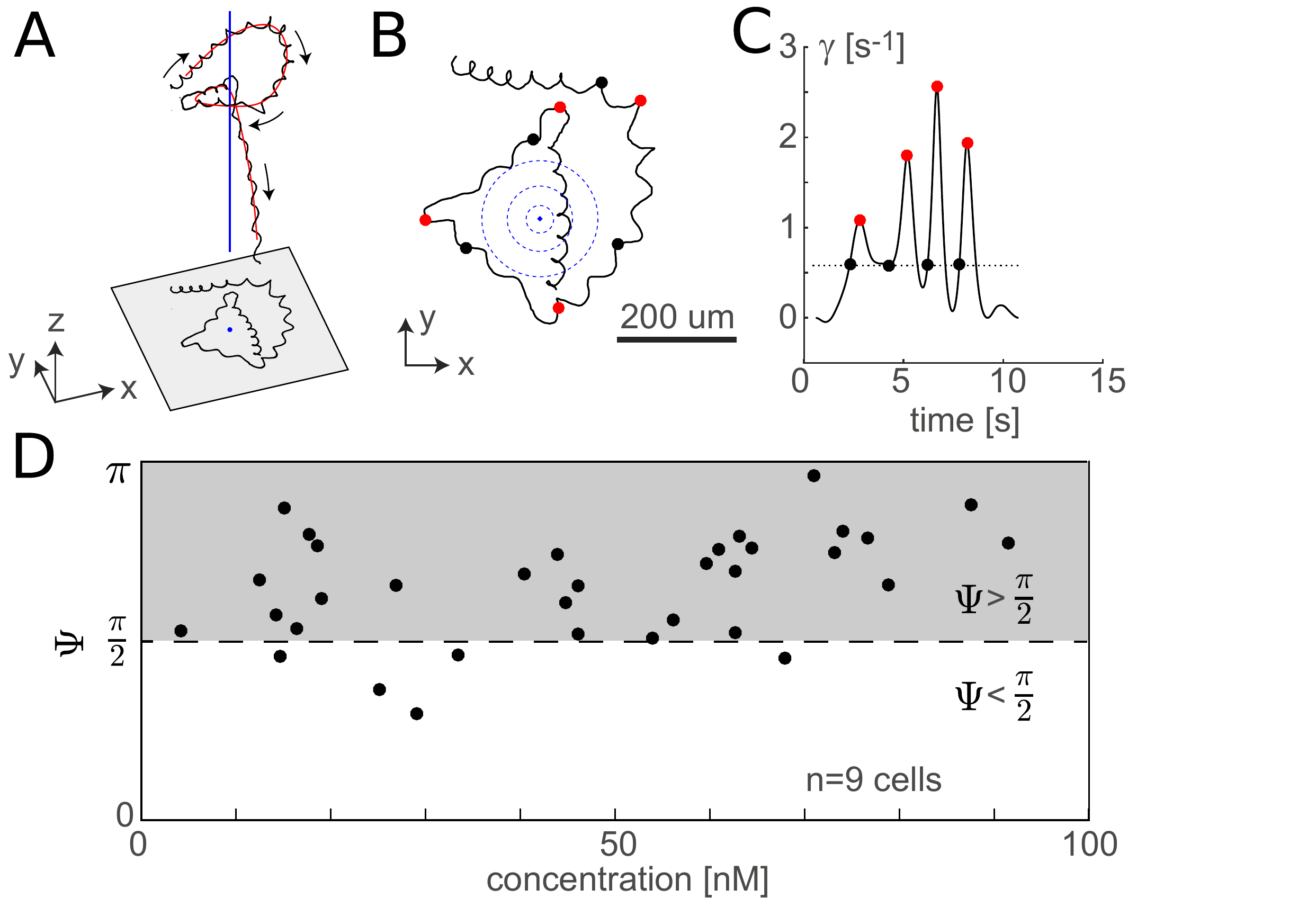}
\caption{
\textbf{Decision making in chemotaxis of sea urchin sperm.} 
(A) Helical swimming path of a sea urchin sperm cell (black) with helix centerline (red),
while navigating in a concentration field of the chemoattractant resact \cite{Jikeli2015}. 
The concentration field is cylindrically symmetric with symmetry axis parallel to the $z$-axis (indicated in blue).
(B) Projection of the same swimming path on the $xy$-plane. 
Dots mark the beginning (black) and peak (red) 
of `high-gain' steering phases (or off-responses \cite{Jikeli2015}).
The concentration field is indicated by blue circles. 
(C) From the swimming path and the local gradient direction, 
we can determine a time-dependent rate $\gamma(t)$ of helix bending towards the gradient \cite{Jikeli2015}.
The beginning of a `high-gain' steering phase is defined as the level-crossing of $\gamma(t)$ 
above its median as is indicated by black dots. Peaks of $\gamma(t)$ are indicated by red dots.
(D) Scatter plot of the orientation angle $\Psi$ and local concentration $c$ at the beginning of `high-gain' steering phases ($n=9$ cells). 
`High-gain' steering is predominantly initiated for $\Psi>\pi/2$ (gray shading).}
\label{fig:exp}
\end{figure}

Recent experiments revealed that during their chemotaxis along helical paths, 
sea urchin sperm cells switch between 
two distinct steering modes in a situation-specific manner \cite{Jikeli2015}, see Fig.~\ref{fig:exp}.
These two steering modes, termed on- and off-response, are characterized by low and high values of the 
the rate $\gamma$ of helix bending in the direction of the local concentration gradient, 
respectively, see Fig.~\ref{fig:exp}C. 
Cells were observed to employ on-responses when their helix axis pointed in the direction of the concentration gradient,
but initiated a transient off-response, if their helix axis pointed down the gradient \cite{Jikeli2015}, see Fig.~\ref{fig:exp}D. 
Here, we defined the start of an off-response with `high-gain' steering
as the level crossing of $\gamma$ above its median,
and recorded the angle $\Psi$ between the current swimming direction and local concentration gradient at the respective times,
see Supporting Material (SM) for details.
The dynamic regulation of the bending rate $\gamma$ as a function of the orientation angle $\Psi$
can be considered as a dynamic switching between two steering modes.
Note that this relationship cannot be explained by the simple geometric relation $\gamma\sim\sin\Psi$ predicted by a previous theory 
\cite{Friedrich2007}, see Fig.~S1 in SM text. 

The benefit of dynamic switching between steering modes is not known.
Our theory provides a strong rationale that this dynamic switching 
increases the probability to find the egg in the presence of noise. 

\section{Theory of helical chemotaxis}

\subsection{Swimming, signaling, and steering}

We consider a theoretical description of sperm chemotaxis along helical paths,
which describes the feedback loop between swimming, chemotactic signaling, and steering \cite{Friedrich2007,Friedrich2009}.
We extend this theory by incorporating a situation-specific modulation of the sensori-motor gain factor,
which can take two different values in our theory.
This represents a simple implementation of decision making.

During chemotactic navigation, 
a sperm cell measures the concentration of chemoattractant along its swimming path $\textbf{r}(t)$.
At low concentrations, 
the rate $b(t)$ at which chemoattractant molecules bind to receptors on the cellular membrane is proportional to the local concentration $c(\textbf{r}(t))$, {i.e.}
\begin{eqnarray}
\label{eq:RateResactBinding}
b(t)=\lambda\, c(\textbf{r}(t))
\end{eqnarray}
with binding constant $\lambda$ \cite{Pichlo2014}. 
The input $s(t)$ to the chemotactic signaling system is given by 
the train of individual binding events with rate $b(t)$ (which represents an inhomogeneous Poisson process with arrival times $t_j$)
\begin{align}
\label{eq:MolecularShotNoise}
s(t)=\sum \limits_{j} \delta(t-t_j), \ \ \ \langle s(t) \rangle = b(t),
\end{align}
see Fig.~\ref{fig:figure1} A and B.
We employ a minimal description of chemotactic signaling 
with a dimensionless output of the signaling system $a(t)$ and a dynamic sensitivity $p(t)$ \cite{Friedrich2009}, 
which implements its main characteristics: sensory adaptation and relaxation \cite{Kashikar2012}
\begin{eqnarray}
\label{eq:signalingSystem}
\mu\, \dot{a} & = & p[\lambda c_{b}+s(t)]-a,\nonumber \\
\mu\, \dot{p} & = & p(1-a).
\end{eqnarray}
Here, $c_b$ sets a threshold of sensory adaption and $\mu$ characterizes a time scale of relaxation and adaptation. 
Dots denote time derivatives.
For oscillatory input, $s(t)=s_0+s_1\cos(\Omega t)$, 
the output $a(t)$ oscillates around its steady-state value $1$ with amplitude proportional to $s_1/(\lambda c_b+s_0)$.

For chemotactic steering, the output of the signaling system, $a(t)$, dynamically regulates 
the curvature $\kappa(t)$ and torsion $\tau(t)$ of the helical swimming path
\begin{eqnarray}
\label{eq:CouplingToCurvature}
\kappa(t) & = & \kappa_{0}-\rho\,\kappa_0(a-1),\nonumber \\
\tau(t) & = & \tau_{0}+\rho\,\tau_0(a-1).
\end{eqnarray}
Curvature and torsion uniquely characterize the time evolution of the swimming path $\mathbf{r}(t)$ 
by the Frenet-Serret equations, see SM text.
For constant path curvature and torsion, $\kappa(t)=\kappa_0$ and $\tau(t)=\tau_0$, 
the swimming path would be a perfect helix with 
radius $r_0=\kappa_0/(\kappa_0^2+\tau_0^2)$, 
pitch $2 \pi h_0=\tau_0/(\kappa_0^2+\tau_0^2)$, 
and angular helix frequency $\Omega_0=v[\kappa_0^2+\tau_0^2]^{1/2}$, 
where $v$ denotes a constant swimming speed. 
In a concentration field, both $\kappa$ and $\tau$ are dynamically regulated in response to the stochastic input signal $s(t)$. 
The sensori-motor gain factor $\rho$ in Eq.~\ref{eq:CouplingToCurvature}
sets both the speed of chemotactic steering and of noise amplification, 
and will be at our focus in the following.

The chemotaxis paradigm embodied in Eq.~\ref{eq:RateResactBinding}-\ref{eq:CouplingToCurvature} is summarized in Fig.~\ref{fig:figure1} C and D:
Helical swimming around a centerline $\textbf{R}$ with helix axis perpendicular to a concentration gradient $\nabla c$ 
results in oscillations of the binding rate $b(t)$ with the frequency $\Omega_0$ of helical swimming. 
As a consequence, path curvature and torsion oscillate, causing the helix to bend in the direction of the gradient. 
This decreases the angle $\Psi$ between the helix axis and the gradient direction. 
Molecular shot noise in concentration measurements adds stochasticity to this directed steering, as discussed next.


\subsection{High-gain steering amplifies sensing noise}

Eqs.~\ref{eq:RateResactBinding}-\ref{eq:CouplingToCurvature} (with Eqs.~S3-S5 in SM) represent a closed control loop and can be simulated numerically to obtain sperm swimming paths. 
We use a representative concentration field $c(\mathbf{x})$, established by diffusion from a spherical source representing an egg. 
Parameters have been chosen to match experiment, see SM text.
In particular, we use measured values for the chemoattractant content of egg cells and the diffusion coefficient of the  chemoattractant \cite{Kashikar2012}.
Thus, computed concentrations and corresponding noise levels are representative of physiological conditions in sea urchin. 

\begin{figure*}[thb]
\begin{center}
\includegraphics[width=0.8\linewidth]{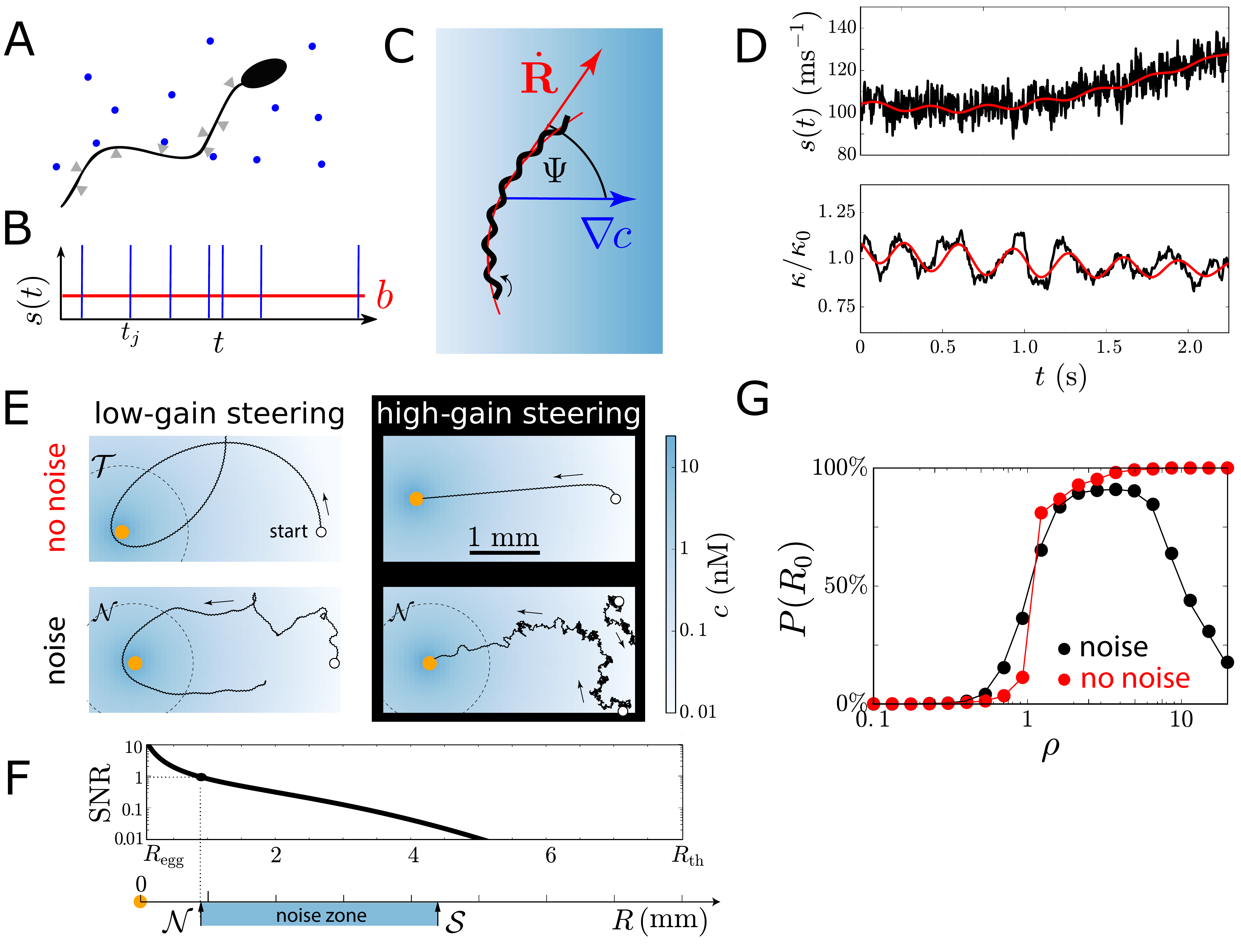}
\end{center}
\caption{
\textbf{Helical chemotaxis in the presence of sensing noise.} 
(A) Chemoattractant molecules bind to receptors on the cell membrane. 
(B) The sequence of binding events defines a stochastic input signal $s(t)$ with rate $b(t)$, Eq.~\ref{eq:RateResactBinding}.
(C) A sperm cell swims along a helical swimming path (black),
whose centerline (red) can bend in the direction of a concentration gradient (blue).
(D) Helical swimming in a concentration gradient causes a periodic modulation of the rate $b(t)$ of binding events (red). 
Representative realization of input signal $s(t)$ (black, low-pass filtered for visualization).
This signal dynamically regulates the path curvature $\kappa(t)$, 
here shown in the absence of sensing noise (red) and for stochastic input signal (black).
(E) 
Example swimming paths with and without sensing noise
for two values of the gain factor 
(`low-gain' steering $\rho_\mathrm{low}=1$, `high-gain steering' $\rho_\mathrm{high}=10$).
Egg cell (yellow disk).  
(F)
signal-to-noise ratio (SNR) as a function of distance $R$ from the egg.
The SNR defines a `noise zone' spanning intermediate distances $R$,
bounded by a noise zone boundary $\mathcal{N}$, where $\mathrm{SNR}=1$,
and a spatial limit of chemosensation $\mathcal{S}$, where $c(R)=(\lambda T)^{-1}$.
(G) 
Probability to find the egg as a function of gain factor $\rho$ 
for initial distance $R_0=3\ \mathrm{mm}$ to the egg (and random initial orientation). 
Without sensing noise, the success probability increases monotonically with $\rho$, 
while in the presence of noise, this probability displays a maximum at an optimal $\rho$. 
Maximum search time $300\,\mathrm{s}$. Error bars smaller than symbols.
Parameters, see SM text.
}
\label{fig:figure1} 
\end{figure*}

Fig.~\ref{fig:figure1}E shows swimming paths both in the absence and presence of sensing noise,
for a low and a high value of the gain factor $\rho$ in Eq.~\ref{eq:CouplingToCurvature}, respectively. 
For `low-gain' steering, and in the absence of noise, {i.e.} $s(t)=b(t)$,
the model sperm cell moves closer to the egg at first, but misses the egg.
In fact, the same occurs for all initial conditions with egg distance $R_0=|\textbf{R}(t=0)|$ 
in a range $\mathcal{T}<R_0<\mathcal{A}_\mathrm{low}$,
with $\mathcal{T}\approx 1.0\,\mathrm{mm}$ and $\mathcal{A}_\mathrm{low}\approx 3.8\,\mathrm{mm}$
if the helix axis is initially perpendicular to the gradient direction, 
see SM text.
For initial distances outside this attraction zone, $R_0>\mathcal{A}_\mathrm{low}$,
swimming paths move away from the egg due to insufficient chemotactic attraction.
In a `target zone' defined by $R_0<\mathcal{T}$,
the direction of the concentration gradient changes on short length scales due to the radial symmetry of the concentration field,
and helix bending during `low-gain' steering is too slow to follow the gradient.

In the presence of noise, swimming paths become stochastic.
For `low-gain' steering, with only slight course correction, 
noise in the input signal hardly affects swimming paths. 
In contrast, 
a high gain factor results in fast bending of helical paths, 
yet it amplifies noise in concentration measurements considerably. 
This is particularly evident in a `noise zone' spanning intermediate distances $R$ from the egg, 
where concentration signals are detectable, but the signal-to-noise ratio (SNR) of gradient measurements is below one,
see Fig.~\ref{fig:figure1}F.
We define the SNR as the ratio between 
the power of the gradient signal (here encoded in oscillations of the binding rate $b(t)$ 
with amplitude $\lambda|\nabla c|r_0$ for swimming perpendicular to the gradient direction),
and the noise strength of the input signal $s(t)$ relative to a single helix period of duration $T$
\begin{eqnarray}
\label{eq:SNR}
\text{SNR}(R) =
\frac{(\lambda |\nabla c| r_0)^2/2}{\lambda c_0/T},
\end{eqnarray}
see SM text for a derivation.
We introduce the distance $\mathcal{N}$ where the SNR equals one.
Additionally, we introduce a distance $\mathcal{S}$ where only one molecule will be detected per helical turn on average, 
which marks a spatial limit of chemosensation.
These two distances provide a formal definition of the `noise zone' as the range of distances 
$\mathcal{S}<R<\mathcal{N}$
bounded by $\mathcal{N}$ and $\mathcal{S}$.


\subsection{Optimal chemotactic gain factor}

We computed the probability $P(R_0)$ to find the egg for a given initial distance $R_0$ from the egg
for a static gain factor $\rho$ in Eq.~\ref{eq:CouplingToCurvature}, 
see Fig.~\ref{fig:figure1}G
(assuming an isotropic distribution of initial swimming directions).
In the hypothetical case of noise-free concentration measurements, 
the success probability is a monotonically increasing function of $\rho$.
For physiological levels of sensing noise, however,
we predict an optimal value of the gain factor $\rho$ that maximizes $P(R_0)$, 
reflecting the competition between responding accurately or responding fast. 


\subsection{Speed of steering and directional fluctuations are inseparably coupled}

The centerline $\R$ of helical paths describes a stochastic trajectory with directional persistence.
The dynamics of its tangent vector $\dot{\R}/|\dot{\R}|$ can be formally described 
as a superposition of
(i) bending in the direction of the local concentration gradient with bending rate $\gamma$, and 
(ii) effective rotational diffusion with rotational diffusion coefficient $D$, see SM text. 
The bending rate $\gamma$ characterizes a noise-averaged steering response, 
corresponding to the expectation value $b(t)$ of the input signal $s(t)$, 
whereas the rotational diffusion coefficient $D$ characterizes directional fluctuations 
of the tangent vector that arise from fluctuations of the input signal around its expectation value.
An analytical theory valid in the limit of weak concentration gradients with $|\nabla c r_0|/c\ll 1$
provides expressions for both $\gamma$ and $D$, 
demonstrating how both quantities scale with the sensori-motor gain factor $\rho$
\begin{align}
\label{eq:gamma}
\gamma &= \frac{\rho \varepsilon}{T} \, \frac{|\nabla_\perp c|}{c_b+c}, \\
\label{eq:D}
D &= \left(\frac{\rho \varepsilon}{T}\right)^2 \, \frac{c}{\lambda(c_b+c)^2}.
\end{align}
Here,
$\varepsilon=2\pi\kappa_0\tau_0/(\kappa_0^2+\tau_0^2)$ 
is a geometric factor characterizing helical swimming.
Eqs.~\ref{eq:gamma} and \ref{eq:D} 
were previously derived for the special case of a linear concentration field 
\cite{Friedrich2007, Friedrich2009} and generalized here.
Note that the effective rotational diffusion coefficient $D$ depends on the concentration $c$ of signaling molecules.

The ratio between the bending rate $\gamma$ and the effective rotational diffusion coefficient $D$
is directly related to the signal-to-noise ratio SNR as defined in Eq.~\ref{eq:SNR}
\begin{equation}
\label{eq:ratio_gamma_D}
\frac{\gamma^2}{D}=\frac{2\sin^2\Psi}{T}\cdot\mathrm{SNR}.
\end{equation}
Eq.~\ref{eq:ratio_gamma_D} 
implies that the speed of steering (characterized by $\gamma$) 
and the strength of directional fluctuations due to sensory noise (characterized by $D$)
are inseparably coupled.

\section{Chemotaxis as a decision problem}

Prompted by recent experiments \cite{Jikeli2015} displayed in Fig.~\ref{fig:exp},
we now address dynamic switching between modes of `low-gain' and `high-gain' steering. 
We consider sperm navigation as a decision problem,
in which a single chemotactic agent, here the sperm cell, can choose between two actions, 
{i.e.} `low-gain' or `high-gain' steering, at each state.
We ask for a strategy that maximizes the probability to find the egg.
To this end, we map the stochastic dynamics of sperm chemotaxis on a finite-state Markov decision process (MDP) \cite{Bellman1957}.

\begin{figure}[t]
\includegraphics[width=\linewidth]{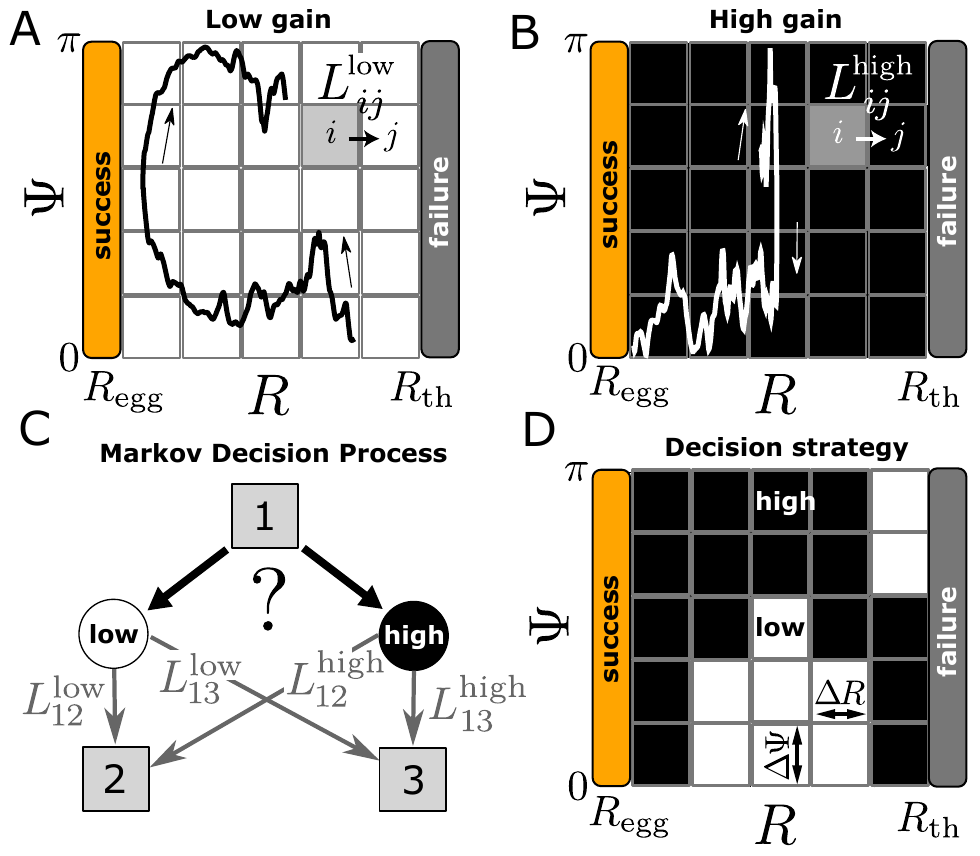}
\caption{
\textbf{Sperm navigation mapped on a Markov decision process.} 
(A,B) Binning of $(R,\Psi)$-phase space and sketch of trajectories for `low-gain' (white) and `high-gain' (black) steering. 
(C) Illustration of a single decision:
Starting in a state $1$, the player first chooses between two actions, {i.e.} `low-gain' steering or `high-gain' steering. 
This choice determines the transition probabilities $L_{ij}$ for jumping to a different state, here labeled 2 and 3.
(D) Illustration of a decision strategy, assigning a choice of action to each state.
The figure shows coarse bins for sake of illustration. 
}
\label{fig:figure2}
\end{figure}

We characterize simulated swimming paths by their time-dependent distance $R(t)=|\textbf{R}(t)|$ to the egg, 
and a time-dependent orientation angle $\Psi(t)$, as defined in Fig.~\ref{fig:figure1}C.
Symmetry implies that the two variables, $R$ and $\Psi$, 
are sufficient to describe the full dynamics of steering 
in a coarse-grained, analytical theory of sperm chemotaxis \cite{Friedrich2007}.
We first discretize $(R,\Psi)$-phase space, mapping simulated swimming paths
(corresponding to Eq.~\ref{eq:RateResactBinding}-\ref{eq:CouplingToCurvature} for static $\rho$),
on a sequence of discrete states, see Fig.~\ref{fig:figure2}.
Using ensembles of $10^4$ swimming paths, we determine transition probabilities
for the transition from one bin labeled $i$ to another bin $j$.
This is done for two values of the gain factor, $\rho=\rho_\mathrm{low}$ or $\rho=\rho_\mathrm{high}$, 
yielding respective transition matrices $L_{ij}^\mathrm{low}$ and $L_{ij}^\mathrm{high}$.
In each case, the transition dynamics is approximately Markovian, see Fig.~S5 in SM text. 
We additionally introduce an absorbing `success state', if the egg is found, and an absorbing `failure state', 
if the cell moves beyond a threshold distance, 
marking the end of a single search attempt.
This defines a Markov chain, which allows us to efficiently determine probabilities to eventually reach the `success state'.
As a control, we compare success probabilities computed using full simulations with static gain factor and predictions from this Markov chain, see Fig.~S6 in SM text. 

Now, we consider the corresponding MDP, 
where the model cell can choose between the two actions `low-gain' steering and `high-gain' steering in each state. 
This choice determines the transition probabilities $L_{ij}$ to the next state, 
which are taken as either $L_{ij}^\mathrm{low}$ or $L_{ij}^\mathrm{high}$ depending on the action chosen, 
see Fig.~\ref{fig:figure2}C for illustration.
We can ask for the optimal strategy, {i.e.}, a rule which action to choose in a given state at a given time
in order to maximize the probability to eventually reach the `success state'.
An example strategy is sketched in Fig.~2D, assigning a choice of action to each state.
A fundamental theorem in the theory of MDP states that a strategy with fixed choice for each state, 
independent of the history of previous states, is sufficient to maximize the probability to reach the `success state'~\cite{Baier2008}.
We now compute such optimal strategies, and discuss how these depend on the presence of sensing noise.

\subsection{Decision making increases success probability}

We computed optimal, memoryless decision strategies for the MDP of sperm navigation,
using the probabilistic model checker PRISM~\cite{Kwiatkowska2011}, see SM for details. 
In Fig.~\ref{fig:figure3}, we compare the success probability for the optimal strategy to the success probabilities 
one would obtain for strategies that choose either always `low-gain' or `high-gain' steering.
In the hypothetical case of noise-free concentration measurements, 
the performance of the optimal strategy is virtually indistinguishable to that of `high-gain' steering, see Fig.~\ref{fig:figure3}A. 
In contrast, when accounting for physiological levels of sensing noise, 
success probabilities for the optimal strategy are substantially higher than success probabilities for `low-gain' and `high-gain' steering, see Fig.~\ref{fig:figure3}B.
This concerns especially intermediate initial distances from the egg, 
where concentrations are rather low and sensing noise corrupts concentration measurements. 

\subsection{Decision making important in noise zone} 
\begin{figure}[t]
\begin{minipage}{\linewidth}
\includegraphics[width=1\linewidth]{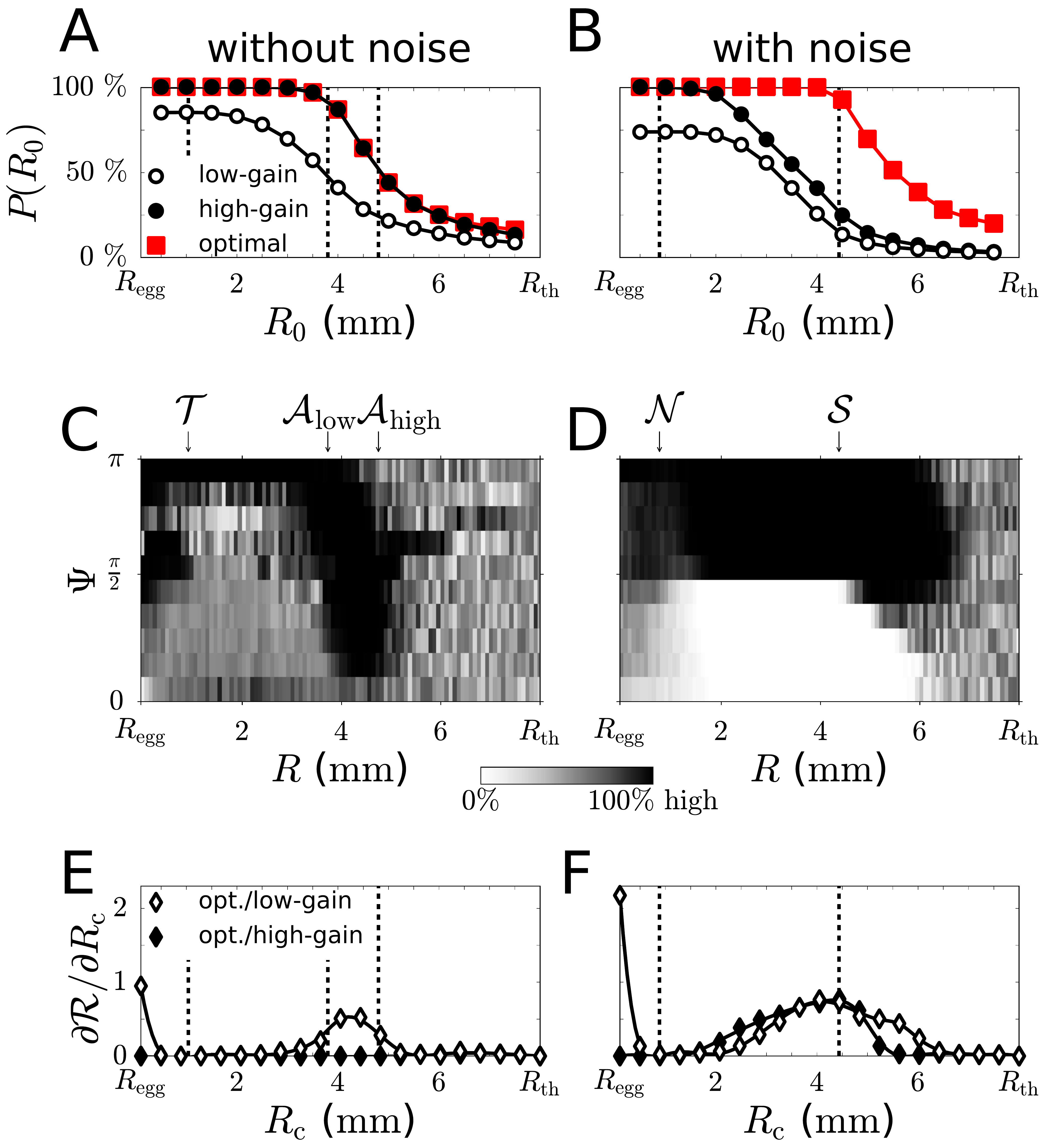}
\end{minipage}
\caption{
\textbf{Chemotactic success with decision making.}
Success probability $P(R_0)$ for the optimal decision strategy, 
resulting from switching between `low-gain' and `high-gain' steering 
as function of initial distance $R_0$ to the egg 
for the case of noise-free concentration measurements (A), and physiological levels of sensing noise (B) (red squares). 
For comparison, success probabilities for strategies without decision making are shown (circles). 
(C,D) Optimal decision strategies for the cases shown in panel A and B.
Grayscale represents prediction frequency of `high-gain' steering,
using a cohort of MDPs obtained by bootstrapping, see SM text for details.
Arrows and dashed lines indicate zone boundaries as introduced in Fig.~\ref{fig:figure1}.
(E,F) 
Spatial sensitivity analysis of optimal strategies:
Shown is the change in chemotactic range $\mathcal{R}$ as function of cut-off distance $R_c$ for hybrid strategies 
that employ the optimal strategy for $R<R_c$, and either `low-gain' steering (white circles) or `high-gain' steering (black circles) else.
Positive values indicate a benefit of decision making at the respective distance to the egg.
Parameters, see SM text. 
}
\label{fig:figure3}
\end{figure}

Next, we analyzed significance and benefit of optimal decision making at different distances from the egg.
We averaged computed strategies for an ensemble of realizations of the MDP, 
each with transition probabilities obtained by bootstrapping from a large cohort of simulated sperm swimming paths, 
see Fig.~\ref{fig:figure3} C and D.
Gray-scale values indicate the frequency that `high-gain' steering is predicted to be optimal for a given state.
Thereby, we explicitly harness numerical variations in transition probabilities
to extract relevant features of optimal decision strategies.

For the case of noise-free concentration measurements, 
we find two distinct state-space regions, where `high-gain' steering is always favored.
The first region corresponds to the `target zone', defined as $R<\mathcal{T}$, 
where the model sperm cell cannot come closer to the egg 
if it employs `low-gain' steering and initially starts with its helix axis perpendicular to the concentration gradient.
A second region is bounded from below by the attraction radius 
$\mathcal{A}_\mathrm{low}\approx 3.8\,\mathrm{mm}$
for `low-gain' steering,
and an analogously defined $\mathcal{A}_\mathrm{high}$ for `high-gain' steering
with $\mathcal{A}_\mathrm{high}\approx 4.8\,\mathrm{mm}$. 
In this region, `high-gain' steering is important to attract cells that would otherwise move away from the egg, see SM text.

In the presence of sensing noise,
we consistently find that the optimal strategy chooses `low-gain' steering while moving up-gradient, 
but chooses `high-gain' steering when accidentally moving down-gradient, see Fig.~\ref{fig:figure3}E. 
This choice is prevalent in the `noise zone' at intermediate distances from the egg, 
where gradients are detectable, but the SNR ratio is below one. 
There, `low-gain' steering minimizes the amplification of sensing noise, 
whereas `high-gain' steering allows for fast reorientation responses at the risk of steering in a wrong direction. 
Remarkably, this prediction matches the observed steering behavior of sperm cells 
observed in recent experiments \cite{Jikeli2015}, see Fig.~\ref{fig:exp}D.

To compare the efficacy of different decision strategies, we introduce the effective chemotactic volume 
\begin{eqnarray}
\frac{4}{3}\pi\,\mathcal{R}^3 = \int \limits_0^\infty dR\, (4\pi R^2) P(R),
\end{eqnarray}
which defines an \textit{effective chemotactic range} $\mathcal{R}$ for a given decision strategy. 
We find $\mathcal{R} \approx 6.2$ $\text{mm}$ for the optimal strategy, 
while $\mathcal{R}_\mathrm{low} \approx 3.8$ $\text{mm}$ and $\mathcal{R}_\mathrm{high} \approx 4.3$ $\text{mm}$ 
for a strategy that always chooses either `low-gain' or `high-gain' steering, respectively. 

Next, we asked at which distances to the target decision making is most important.
To quantify respective benefits, we computed chemotactic ranges $\mathcal{R}(R_c)$ as a function of a cut-off distance $R_c$ 
for hybrid strategies. These hybrid strategies employ the optimal decision strategy only at distances smaller than $R_c$, 
but choose always either `low-gain' or `high-gain' steering, respectively, outside this range. In particular, 
positive values of the derivative $\partial \mathcal{R}/\partial R_c$ reveal 
at which distances decision making is most beneficial, see Fig.~\ref{fig:figure3} E and F. 
Distances where this spatial significance measure is positive
match exactly those regions where decision strategies are most stable with respect to numerical noise.
Thus, two independent spatially-resolved sensitivity measures for optimal strategies give congruent results.


\section{Cellular implementation of decision making}

\begin{figure}[t]
\includegraphics[width=1.05\linewidth]{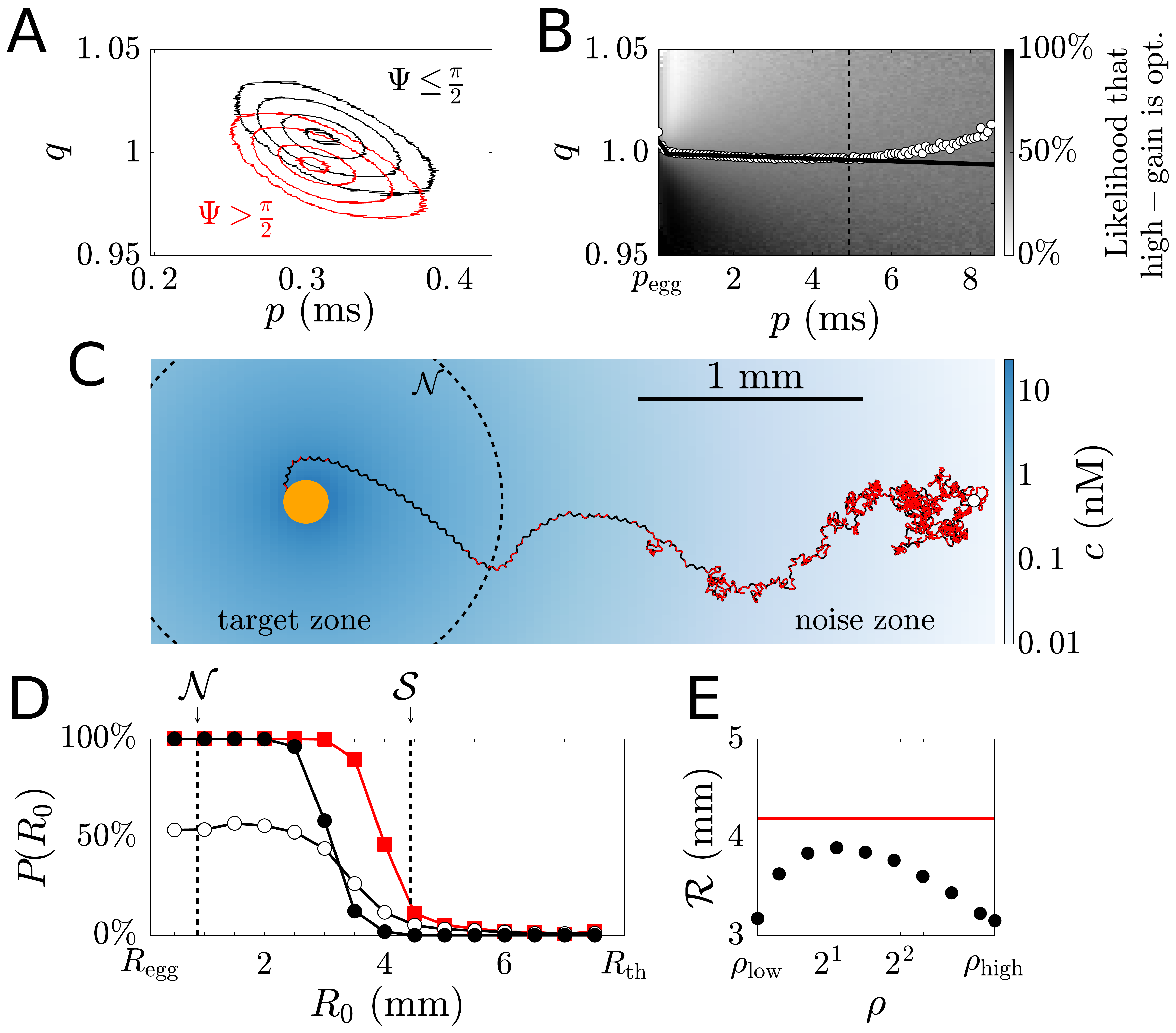}
\caption{\textbf{
Simple implementation of optimal decision making.
} 
(A) signaling variables $p$ and $q$ contain information about the helix orientation angle $\Psi$ and distance $R$ to the target. 
Contour levels for conditional probability densities $P(p,q|R, \Psi > \frac{\pi}{2})$ (red) and $P(p,q|R, \Psi \le \frac{\pi}{2})$ (black) 
(corresponding to 1\%, 10\%, 50\%, 90\% percentiles; $R=1.5\,\text{mm}$).
(B)
Relative frequency of `high-gain' steering predicted by the optimal decision strategy, 
for given combination of $(p,q)$.
We define a decision boundary by a piecewise linear fit to the $50\%$-contour line
(up to $p=5\,\mathrm{ms}$, corresponding to a limit of sufficiently reliable state estimation, see SM text). 
(C) simulated swimming path using this decision rule, 
with dynamic switching between `high-gain' steering (red) and `low-gain' steering (black);
projected on $xy$-plane.
The chemoattractant concentration in this plane is shown (blue gradient), together with the boundary of the noise zone.
(D) Success probability $P(R_0)$ for full simulations with simple decision making (red) 
as a function of initial distance $R_0$ to the egg.
For comparison, success probabilities for `low-gain' steering (white) and `high-gain' steering (black) are shown. 
(E) The effective chemotactic range $R$ with decision making (red) is larger than 
$R$ for an optimal constant gain factor (black). 
Parameters, see SM text.}
\label{fig:figure4}
\end{figure}

While the formalism of MDPs allows us to efficiently compute optimal decision strategies, 
it is not evident how a biological cell would implement such strategies.
In particular, a swimming cell has no direct access to the state variables $R$ and $\Psi$, 
but only to the noisy concentration signal $s(t)$.
We present a minimal signaling system that implements decision making on the basis of $s(t)$, i.e. on the basis of available information.
We introduce a trigger variable $q(t)$ that tracks the output of the signaling system $a(t)$ with a relaxation time scale $\eta$
\cite{Jikeli2015}
\begin{eqnarray}
\label{eq:EstimatePsi}
\eta \dot{q}= a - q.
\end{eqnarray}
This low-pass filter attenuates fast oscillations of $a(t)$ caused by helical swimming in a concentration gradient, 
yet faithfully retains changes in the baseline of $a(t)$, 
which occur for either up-gradient or down-gradient swimming as a result of a finite time scale of sensory adaptation.
In the absence of noise, and for a given concentration field $c(R)$, 
the signaling variables $(p,q)$ are directly related to the state variables $(R,\psi)$ as
$p^{-1}=\lambda[c_b+c(R)]$ and
$q=1+\mu\omega_0h_0 p \lambda |d c(R)/dR| \cos\psi$, 
(if we neglect residual oscillations of $q(t)$).
In the presence of noise, 
$p$ and $q$ scatter around their expected values, see Fig.~\ref{fig:figure4}A.
Consequently, estimation of state $(R,\psi)$ based on $(p,q)$ is associated with an error.
The accuracy in discriminating between swimming up-gradient ($\psi \leq \pi/2$) 
and down-gradient ($\psi > \pi/2$) decreases as a function 
of distance $R$ from the egg, see Fig.~S7 in SM text. 
Estimation of helix orientation angle $\Psi$ can be considered feasible up to a maximal distance $R\approx 3\,\text{mm}$, 
where the accuracy equals $66\%$ ($100\%$: perfect discriminability, $50\%$: complete lack of discriminability).

We now design a decision rule in terms of $p$ and $q$
\begin{equation}
\rho(p,q)=
\begin{cases}
\rho_\mathrm{low},  & \text{ for } q\ge \Theta(p) \\
\rho_\mathrm{high}, & \text{ for } q < \Theta(p)
\end{cases},
\end{equation}
with decision boundary $\Theta(p)$ yet to be determined.
From the optimal decision strategy predicted for the MDP,
we compute the relative frequency of `high-gain' steering for each pair of values $p$ and $q$, 
using the likelihood of states $(R,\Psi)$ for given tuple $(p,q)$, see Fig.~\ref{fig:figure4}B. 
We define $\Theta(p)$ as a piecewise linear fit to the $50\%$-contour line of this relative frequency, see SM text. 
Note that this decision boundary implies `low-gain' steering far from the egg.

Using full simulations of sperm swimming paths with this decision rule, 
we find that decision making indeed increases the probability of success for intermediate initial distances to the egg, similar to our analysis of the MDP, see Fig.~\ref{fig:figure4}D.
Note that we use a finite search time of $300\,\mathrm{s}$ in Fig.~\ref{fig:figure4}D, 
which yields lower success probabilities as the corresponding MDP representation,
which considers infinite search times.
Simulations with longer search times confirm the benefit of cellular decision making for sperm chemotaxis,
see Fig.~S6 in SM text. 
Our simple implementation of decision making is more effective than any constant gain factor, see Fig.~\ref{fig:figure4}E.
While we compute $\Theta(p)$ for a specific concentration field, 
the same decision boundary performs superior also in other concentration fields,
highlighting a general benefit of decision making, see Fig.~S8 in SM text. 

\section{Discussion}





We developed a theory of optimal chemotaxis towards a single target in the presence of noise,
using sperm chemotaxis along helical paths as application example.

We show that a situation-specific switching between two different steering modes 
- `low-gain' and `high-gain' steering -
maximizes the probability to find a target, such as an egg,
at the center of a radial concentration field of signaling molecules.
In the optimal strategy, `low-gain' steering is chosen 
if the cell is approximately heading in target direction. 
This minimizes the risk of inadvertently steering in the wrong direction by 
amplifying noise in the chemotactic input signal.
If, however, the net swimming direction is at least perpendicular to the target direction, 
the potential benefit of fast steering outweighs the risk of wrong course corrections,
and `high-gain' steering is chosen. 

The optimal strategy predicted by our theory 
matches a surprising experimental observation recently made by Jikeli {et al.} \cite{Jikeli2015},
summarized in Fig.~\ref{fig:exp}.
There, it was observed that sea urchin sperm cells switch between `low-gain' and `high-gain' steering
depending on their net swimming direction relative to the local concentration gradient.
Our theory provides a rational why switching between steering modes is beneficial. 
%
The benefit of decision making is causally related to noise in sensory input. 
If cells could measure concentrations with perfect accuracy, decision making would provide no benefit. 
In the presence of noise, however, 
decision making increases the probability to find the egg.
We compute physiological signal-to-noise ratios relevant for sperm chemotaxis, 
and show that noise in cellular gradient measurements poses a key constraint on chemotactic navigation.
Concomitantly, decision making is most important in a noise zone at intermediate distances from the egg, 
where chemical signals are detectable, but signal-to-noise ratios of gradient measurements are below one.
Cells must cross this noise zone when arriving from a distance to the target.

The centerline of helical swimming paths represents the trajectory of a persistent random walker,
whose swimming direction vector continuously rotates in the direction of 
the local gradient of the concentration field,
while being subject to directional fluctuations.
Thus, sperm chemotaxis along helical paths provides an example of a biased persistent random walk. 
In this generic picture, decision making amounts to a situation-specific regulation 
of the persistence length of the effective swimming path.  
The adaptive control of persistence length represents 
a time-continuous variant of the `run-and-tumble' chemotaxis strategy employed 
by bacteria such as \textit{E.~coli} \cite{sourjik2012responding}.
The time-continuous variant as presented here has, 
to the best of our knowledge, not been described before,
and allows cells to harness inevitable noise for purposeful navigation. 

Our work reveals a fundamental relationship between the 
speed of chemotactic steering and the strength of directional fluctuations, 
which are caused by noise in the sensory input.
The resultant trade-off between either reliable or fast steering
applies to chemotactic motion with directional persistence in general. 
In particular, Eqs.~\ref{eq:gamma} and \ref{eq:D} will hold in similar form for a chemotaxis strategy of spatial comparison
with spatially extended sensor arrays as employed {e.g.} by eukaryotic cells with crawling motility. 
We thus anticipate that our new approach of mapping 
chemotaxis in the presence of noise on a Markov decision process 
can be generalized to compute optimal navigation strategies for a variety of other search problems.

In conclusion, we propose that decision making between different steering modes
represents a versatile strategy for navigation in the presence of strong noise. 
The optimal strategy predicted by our theory requires only minimal computational capacities of 
chemotactic agents and could inspire optimal control designs for artificial microswimmers.

\begin{acknowledgments}
The authors are supported by the DFG through the Excellence Initiative by the German Federal and State Governments (cluster of excellence cfaed).
We thank U.B.~Kaupp, L.~Alvarez, and J. Karschau and all other  members of the Biological Algorithms Group for stimulating discussions.
Experimental data analyzed in Fig.~\ref{fig:exp} was kindly provided by J.~Jikeli \cite{Jikeli2015}.
\end{acknowledgments}

\nocite{Farley2001,AitSahalia2010,Bellman1957,Puterman1994,Haddad2014,Baier2017}

\newpage

\end{document}